**Magnetic control of Goos-Hänchen shifts in a yttrium-iron-garnet film**

Wenjing Yu,[1] Hua Sun,[1,3] & Lei Gao[1,2,*]

[1]College of Physics, Optoelectronics and Energy of Soochow University, Collaborative Innovation Center of Suzhou Nano Science and Technology, Soochow University, Suzhou 215006, China.

[2]Jiangsu Key Laboratory of Thin Films, Soochow University, Suzhou 215006, China.

[3]hsun@suda.edu.cn

*leigao@suda.edu.cn

**Abstract:**

We investigate the Goos-Hänchen (GH) shifts reflected and transmitted by a yttrium-iron-garnet (YIG) film for both normal and oblique incidence. It is found that the nonreciprocity effect of the MO material does not only result in a nonvanishing reflected shift at normal incidence, but also leads to a slab-thickness-independent term which breaks the symmetry between the reflected and transmitted shifts at oblique incidence. The asymptotic behaviors of the normal-incidence reflected shift are obtained in the vicinity of two characteristic frequencies corresponding to a minimum reflectivity and a total reflection, respectively. Moreover, the coexistence of two types of negative-reflected-shift (NRS) at oblique incidence is discussed. We show that the reversal of the shifts from positive to negative values can be realized by tuning the magnitude of applied magnetic field, the frequency of incident wave and the slab thickness as well as the incident angle. In addition, we further investigate two special cases for practical purposes: the reflected shift with a total reflection and the transmitted shift with a total transmission. Numerical simulations are also performed to verify our analytical results.

**Introduction:**

The GH effect refers to the lateral shift of an incident beam of finite width upon reflection from an interface which was first studied by Goos and Hänchen[1,2] and theoretically explained by Artmann in terms of the stationary-phase approach in the late 1940s[3]. Since then, such effect has been very important with development of the laser beams and integrated optics[4] and has significant impact on applications as well as for investigations of the fundamental problems in physics. And the studies have been extended from a simple dielectric interface to more complex structures or exotic materials such as metal-dielectric nanocomposites[5,6], epsilon-near-zero metamaterials[7], graphene[8-10], PT-symmetric medium[11], topological insulator[12] etc.

    The GH shift by magneto-optical (MO) materials[13-18] is obtained by making use of ferromagnetic resonances of natural magnetic materials. Similar relation between GH effects and intrinsic resonances was also reported in nonmagnetic dielectric, such as GH shifts arising from phonon resonances in crystal quartz[19]. But what was found in MO materials is of particular interest because of the nonreciprocity in scattering

coefficients originated from the broken time reversal symmetry[20]. As a result, a lateral shift for reflection will occur at the interface between the vacuum and an magnetic material arranged in the Voigt geometry even at normal incidence[14,15], with both sign and magnitude controlled by the applied magnetic field. And the polarization-dependence of the GH shift by MO materials makes it possible to separate the incident radiation into beams of different polarizations[21]. However, the details of the magnetic effects on GH shift are stilled obscure. Most studies only discussed the effects of a semi-infinite antiferromagnetic material—$MnF_2$ at low temperature ($T$=4.2K), with a dispersion quite different from that of conventional MO materials adopted in applications. The role of material properties and geometric factors (such as finite slab thickness, incident angles *etc.*) remains unclarified in the magnetic control of GH shifts with MO materials.

Hence we are motivated to perform a theoretical investigation of the GH shifts reflected and transmitted by a MO slab made of yttrium-iron-garnet (YIG). As a ferrite well known for its high MO efficiency and low damping[22-27], YIG has been extensively studied and broadly adopted in microwave[28-30] and magneto-optics technologies[31-33]. The recent realization of one-way waveguides based on YIG photonic crystals sparks even more interest of the application of this traditional MO material in the field of subwavelength optics[34,35]. It was also shown that hyperbolic dispersion and negative refraction initially investigated in antiferromagnetic materials[36] can be extended to and realized in conventional ferrites[37]. But the GH-shift effects due to a surface/slab of YIG have not been studied.

In this paper we present a theoretical analysis of the lateral shifts of both the reflected beam and the transmitted beam due to a magnetized YIG slab in the Voigt geometry. It is shown that the nonreciprocity effect caused by the MO material does not only result in a nonvanishing reflected shift at normal incidence, but also leads to a slab-thickness-independent term which breaks the symmetry between the reflected and transmitted shifts at oblique incidence. The asymptotic behaviors of the normal-incidence reflected shift are obtained in the vicinity of two characteristic frequencies ($\omega_r$ and $\omega_c$) corresponding to a minimum reflectivity and a total reflection, respectively. And the coexistence of two types of negative-reflected-shift (NRS) at oblique incidence is discussed. We also investigate two special cases for practical purposes: the reflected shift with a total reflection and the transmitted shift with a total transmission. Analytical expressions of the shifts in these cases are obtained approximately, which is in good agreement with the results from numerical calculations.

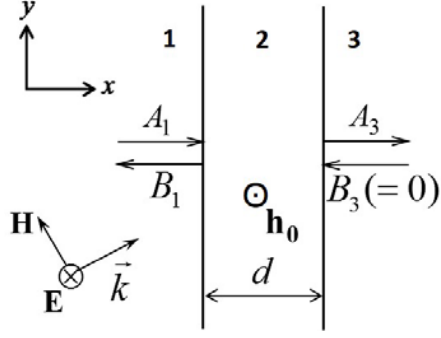

FIG. 1. Schematic diagram of the structure in the presence of an external field $\mathbf{h_0}$. The incident plane wave is polarized along the z-direction and propagates along $\vec{k}$.

## Results

**General formulas.** Consider a YIG film of thickness $d$ surrounded by a non-magnetic background medium of $(\varepsilon_1, \mu_1)$ as shown in Fig. 1. For simplicity we set the background medium in Region 1&3 as the vacuum. The magnetic permeability of YIG magnetized along the z-axis is of the tensor form

$$\ddot{\mu} = \begin{pmatrix} \mu_r & -i\mu_i & 0 \\ i\mu_i & \mu_r & 0 \\ 0 & 0 & 1 \end{pmatrix} \qquad (1)$$

where the digonal and off-diagonal permeabilities follow the typical dispersion of ferrites in the microwave region

$$\mu_r = 1 + \frac{\omega_m(\omega_0 + i\alpha\omega)}{(\omega_0 + i\alpha\omega)^2 - \omega^2} \qquad (2a)$$

$$\mu_i = \frac{\omega_m \omega}{(\omega_0 + i\alpha\omega)^2 - \omega^2} \qquad (2b)$$

Here $\omega$ is the frequency of the incident light, $\omega_0$, $\omega_m$ are the magnetic resonance frequencies given by

$$\omega_0 = 2\pi\gamma h_0, \qquad (3a)$$

$$\omega_m = 2\pi\gamma m_s \qquad (3b)$$

with $h_0$ and $m_s$ denoting the applied magnetic field and the saturated magnetization, respectively. The material parameters of YIG are chosen as: $\gamma = 2.8 \times 10^{-3}$ GHz/Oe, $m_s = 1800$ Gauss, $\varepsilon = 14.5$ [35]. The damping factor $\alpha$ is quite small and neglected in the following analytic derivations and calculations. However, later in the numerical simulations, we have considered the influence of the realistic damping of YIG material.

To find the GH shifts due to such a YIG slab, we start by considering an s-polarized plane wave of angular frequency $\omega$ incident from Region 1 at an angle $\theta$. Then the $x$ component of the wave vectors in layers 1, 3 and 2 are given by

$$k_{1x}, k_{3x} = \left[\frac{\omega^2}{c^2} - k_y^2\right]^{1/2}, \tag{4a}$$

$$k_{2x} = \left[\frac{\omega^2}{c^2}\varepsilon\mu_{eff} - k_y^2\right]^{1/2}, \tag{4b}$$

with

$$k_y = k_0 \sin\theta \tag{5}$$

Here $\varepsilon$ is the dielectric constant of YIG and $\mu_{eff}$ is its effective permeability given by[14]

$$\mu_{eff} = (\mu_r^2 - \mu_i^2)/\mu_r, \tag{6}$$

and $k_0 = \omega/c$ is the wave number of the incident radiation in the background vacuum. *Note that $\mu_{eff}$ is only used to calculate the "effective" wave vector $k_{2x}$ in the MO slab, not to replace the slab by an isotropic one.* Then the electric fields and the magnetic fields in layers 1, 2 and 3 can be expressed as

$$\mathbf{E_1} = (A_1 e^{ik_{1x}x} + B_1 e^{-ik_{1x}x})e^{ik_y y - i\omega t}\mathbf{e_z}$$
$$\mathbf{H_1} = \frac{1}{i\omega\mu_0}\nabla\times\mathbf{E_1} \tag{7}$$

$$\mathbf{E_2} = (A_2 e^{ik_{2x}x} + B_2 e^{-ik_{2x}x})e^{ik_y y - i\omega t}\mathbf{e_z}$$
$$\mathbf{H_2} = \frac{1}{i\omega\mu_0\ddot{\mu}_2}\nabla\times\mathbf{E_2} \tag{8}$$

$$\mathbf{E_3} = (A_3 e^{ik_{3x}(x-d)} + B_3 e^{-ik_{3x}(x-d)})e^{ik_y y - i\omega t}\mathbf{e_z}$$
$$\mathbf{H_3} = \frac{1}{i\omega\mu_0}\nabla\times\mathbf{E_3} \tag{9}$$

Based on the boundary conditions $E_{lz} = E_{(l+1)z}\big|_{x=x_l}$, $H_{ly} = H_{(l+1)y}\big|_{x=x_l}$ ($l = 1, 2$), we obtain the reflection and transmission coefficients as

$$r = \frac{AB^*(2i\sin k_{2x}d)}{|A|^2 e^{ik_{2x}d} - |B|^2 e^{-ik_{2x}d}} \tag{10a}$$

$$t = \frac{-4\mu_{eff} k_{1x} k_{2x}}{|A|^2 e^{ik_{2x}d} - |B|^2 e^{-ik_{2x}d}} \tag{10b}$$

with $A = \mu_{eff} k_{1x} - k_{2x} - igk_y$, $B = \mu_{eff} k_{1x} + k_{2x} + igk_y$ and $g = \mu_i/\mu_r$ is the MO Voigt constant of YIG.

When an electromagnetic beam of finite width illuminates the slab at an incident angle $\theta_0$, the lateral shifts of the reflected and transmitted beams can be obtained by the stationary phase method[3]

$$d_r = -\left.\frac{d\varphi_r}{dk_y}\right|_{k_y=k_y^0} \tag{11a}$$

$$d_t = -\left.\frac{d\varphi_t}{dk_y}\right|_{k_y=k_y^0} \tag{11b}$$

where $k_y^o = k_0 \sin\theta_0$ and $\varphi_r$, $\varphi_t$ are the phase angles of the reflection and transmission coefficients for plane waves, respectively. Note here the lateral shift of the transmitted beam is measured in the same way as that of the reflected beam[38].

For a transparent YIG slab, $\mu_{eff}$ and $k_{2x}$ are both real when the weak absorption of YIG is neglected (i.e. the damping factor $\alpha$ is assumed to be zero). Then the reflected shift derived from Eq. (10)-(11) includes two parts:

$$d_r = -\left.\frac{d\varphi^{(1)}}{dk_y}\right|_{k_y=k_y^0} - \left.\frac{d\varphi^{(2)}}{dk_y}\right|_{k_y=k_y^0} \tag{12}$$

where $\varphi^{(1)} = \text{Arg}(AB^*)$ while $\varphi^{(2)}$ is the phase angle of the complex variable

$$\xi = \frac{1}{|A|^2 e^{ik_{2x}d} - |B|^2 e^{-ik_{2x}d}} \tag{13}$$

The transmitted shift is only determined by the $k_y$-dependence of $\varphi^{(2)}$:

$$d_t = -\left.\frac{d\varphi^{(2)}}{dk_y}\right|_{k_y=k_y^0} \tag{14}$$

When $g = 0$, we have $A = \mu_r k_{1x} - k_{2x} \equiv A_0$, $B = \mu_r k_{1x} + k_{2x} \equiv B_0$. The results of the lateral shifts are reduced to the case of a nonmagneto slab as investigated in Ref. [38]. The first term in Eq. (12) will vanish since $A_0$ and $B_0$ are both real and symmetric reflected and transmitted shifts will appear. Based on the formulas Eq. (12)-(14), we will discuss the behaviors of the shifts at normal incidence and at oblique incidence, respectively, for a MO slab with $g \neq 0$, in the following sections.

**Normal incidence.** When $g \neq 0$, a $\theta$-dependent imaginary part is added to $A$ or $B$ so that

$$A = A_{0e} - igk_y, B = B_{0e} + igk_y \qquad (15)$$

here $A_{0e}$ and $B_{0e}$ are real parameters for an "effective" slab where the MO permeability tensor is replaced by the magnetic-field-controlled scalar $\mu_{eff}$. Since $|A|^2 - |B|^2 = |A_{0e}|^2 - |B_{0e}|^2, |A|^2 + |B|^2 = |A_{0e}|^2 + |B_{0e}|^2 + 2g^2 k_y^2$, the shift term from $\varphi^{(2)}$ is expected to behave like that of the effective slab when the incident angle approaches zero and finally vanishes at normal incidence.

The first term of Eq. (12) is independent of the slab thickness and contributes a non-vanishing reflected shift at normal incidence:

$$d_r^{(n)} = -\frac{d\varphi^{(1)}}{dk_y}\bigg|_{k_y=0} = \frac{\lambda_0 g}{\pi(\mu_{eff} - \varepsilon)} \qquad (16)$$

By combining Eq. (16) with Eq. (2a) and (2b), we obtain the dependence of $d_r^{(n)}$ on frequency and magnetic field in the form

$$d_r^{(n)}(\omega, H) = \frac{\lambda_0}{\pi(\varepsilon-1)} \frac{\omega_m \omega}{\omega^2 - \beta^2 \omega_m^2} \qquad (17)$$

with

$$\beta = \left(H^2 + \frac{\varepsilon-2}{\varepsilon-1}H - \frac{1}{\varepsilon-1}\right)^{1/2} \qquad (18)$$

Here, $H \equiv \frac{\omega_0}{\omega_m} = \frac{h_0}{m_s}$ is a dimensionless magnetic field reduced by the saturated magnetization of the MO slab. In vicinity of the discontinuity point $\omega_c \equiv \beta\omega_m$ (This discontinuity in the frequency spectrum occurs exactly at the reflection minimum, corresponding to $\mu_{eff} = \varepsilon$)[15], the abrupt transition of $d_r^{(n)}$ from negative to positive can be approximated by

$$d_r^{(n)} \approx \frac{\lambda_0}{2\pi\beta(\varepsilon-1)\eta} \qquad (19)$$

where $\eta \equiv \frac{\omega - \omega_c}{\omega_c}$ describes a small deviation from $\omega_c$. Note that the expression of $\varphi^{(1)}$ is identical to that by a semi-infinite MO material in Ref. [14,15]. So Eq. (16)-(19) are also applied to the case of $d \to \infty$, i.e. a semi-infinite YIG interface.

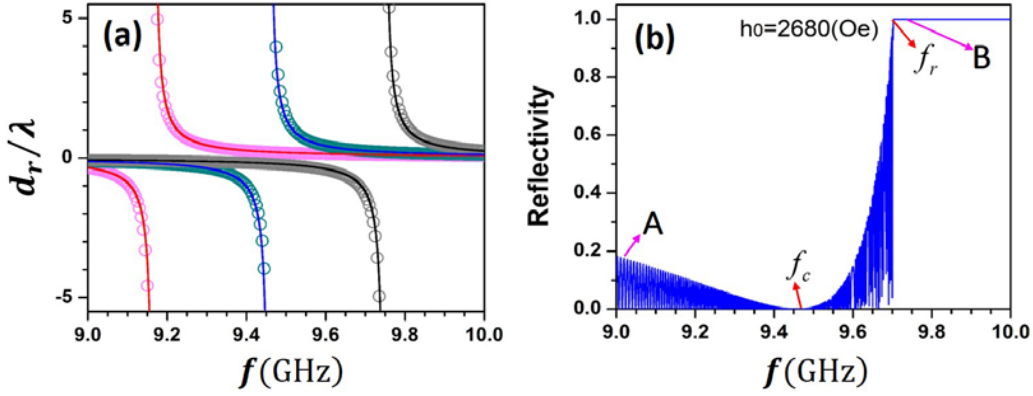

FIG. 2. Calculated normal incidence (a) GH shift of reflected field $d_r/\lambda$ and (b) reflectivity as a function of frequency (express as $\omega/2\pi$). The red, blue and black curves correspond to $h_0 = 2580, 2680, 2780 \text{Oe}$, respectively. Circles: approximated results from Eq. (19); Lines: numerical results.

Fig. 2(a) shows the approximated frequency dependence of $d_r^{(n)}$ based on Eq. (19) for $h_0 = 2580\text{Oe}$, $2680\text{Oe}$ and $2780\text{Oe}$ (circles). The numerical results (lines) directly from Eq. (10) and (11) are displayed simultaneously for comparison and good agreement is found even for moderate deviation from the discontinuity point. Since $\beta$ increases monotonically with $H$, $\omega_c$ is red-shifted when the applied magnetic field $h_0$ is decreased, accompanied by the enhancement of $d_r^{(n)}$ around $\omega_c$. For a lower field $h_0 = 1000\text{Oe}$, we have $d_r^{(n)} \sim \dfrac{0.0136}{\eta}$, which is larger by 1-2 orders of magnitude than the result for MnF$_2$ at the same applied magnetic field as reported in Ref. [14,15].

For practical purposes, a sufficiently large reflectivity is necessary for the application of reflected shift. Fig. 2(b) shows a typical frequency spectrum of reflectivity of a YIG slab ($d = 0.3\text{m}, h_0 = 2680\text{Oe}$), where $|r|^2$ is quite small around $\omega_c$ but rises rapidly when the frequency approaches the sharp edge of a platform of $|r|^2 = 1$. The rapid oscillation of reflectivity is a typical interference pattern of a slab of finite thickness, which is not exhibited in the spectrum of $d_r$ in Fig. 2(a) since the reflected shift is independent of slab thickness. The total-reflection platform at $f > f_r$ occurs when the wave vector $k_{2x}$ in YIG becomes imaginary, which means a negative $\mu_{eff}$ in the cases of normal incidence ($k_y = 0$). According to the dispersion relation of $\mu_i$ and $\mu_r$, it is easy to find

$$\omega_r = \sqrt{H(H+1)}\omega_m \qquad (20)$$

and the frequency dependence of $\mu_{eff}$ and $g$ can be expressed as

$$\mu_{eff} = \frac{\left(\omega_r^2 - \omega^2\right)^2 - \frac{\omega_r^2 \omega^2}{H(H+1)}}{\left(\omega_0^2 - \omega^2\right)\left(\omega_r^2 - \omega^2\right)} \qquad (21a)$$

$$g = \frac{1}{\sqrt{H(H+1)}} \frac{\omega_r \omega}{\omega_r^2 - \omega^2} \qquad (21b)$$

Note that at $\omega = \omega_r$, both $g$ and $\mu_{eff}$ go infinite, but their ratio has a finite value

$$\left.\frac{g}{\mu_{eff}}\right|_{\omega=\omega_r} = \sqrt{\frac{H}{H+1}} \qquad (22)$$

Substituting this in to Eq. (16), we obtain the reflected shift at $\omega_r$

$$d_r^{(n)}(\omega = \omega_r) = \frac{\lambda_0}{\pi} \sqrt{\frac{H}{H+1}} \qquad (23)$$

This result tells us the largest $d_r^{(n)}$ achievable when $|r|^2 = 1$, which increases monotonically with the reduced magnetic field $H$ up to a strong-field limit: $\frac{\lambda_0}{\pi}$.

**Oblique incidence.** When the incident beam is at a certain angle $\theta$, the reflected shift $d_r$ and the transmitted shift $d_t$ caused by a YIG-slab of thickness $d$ can be expressed as

$$d_r(\theta, d) = D(\theta) + \Lambda(\theta, d) \qquad (24a)$$

$$d_t(\theta, d) = \Lambda(\theta, d) \qquad (24b)$$

where the thickness-independent part $D(\theta)$ is according to the first term in Eq. (12), given by

$$D(\theta) = g[F_+(\theta) + F_-(\theta)] \qquad (25)$$

with

$$F_\pm(\theta) = \frac{(\mu_{eff} k_{1x} \pm k_{2x}) + \frac{k_y^2}{k_{1x} k_{2x}}(\mu_{eff} k_{2x} \pm k_{1x})}{(\mu_{eff} k_{1x} \pm k_{2x})^2 + g^2 k_y^2} \qquad (26)$$

The expression of $\Lambda(\theta, d)$ can be obtained from Eq. (13) and (14) as

$$\Lambda(\theta,d) = \frac{dG/dk_y}{1+G^2(k_y)} \tag{27}$$

Here we have introduced a function

$$G(k_y) = \frac{|A|^2+|B|^2}{|A|^2-|B|^2}\tan(k_{2x}d) \tag{28}$$

which can be rewritten as

$$G(k_y) = G_{0e}(k_y)\left[1+M(k_y)\right] \tag{29}$$

where $G_{0e}(k_y)$ is the result of $G(k_y)$ for a slab of scalar permeability $\mu_{eff}$ while $M(k_y)$ gives the correction term caused by the tensor form of the slab permeability:

$$G_{0e}(k_y) = \frac{A_{0e}^2+B_{0e}^2}{A_{0e}^2-B_{0e}^2}\tan(k_{2x}d) \tag{30a}$$

$$M(k_y) = \frac{g^2}{\mu_{eff}^2+1}\frac{k_y^2}{(\mu_{eff}^2+\mu_{eff}\varepsilon)k_0^2-k_y^2} \tag{30b}$$

Note that $g/\mu_{eff} = \frac{\omega_m\omega}{(H+1)^2\omega_m^2-\omega^2}$, hence the condition $g^2/(\mu_{eff}^2+1) \ll 1$ holds for most frequencies not close to $\omega_r$ in the transparent region $\omega < \omega_r$, and $\Lambda(\theta,d)$ can be well approximated by the shifts $d_{re} = d_{te} = \Lambda_e(\theta,d)$ due to an effective non-MO slab of $(\varepsilon, \mu_{eff})$ for the same incident angle $\theta$ and slab-thickness $d$ [38].

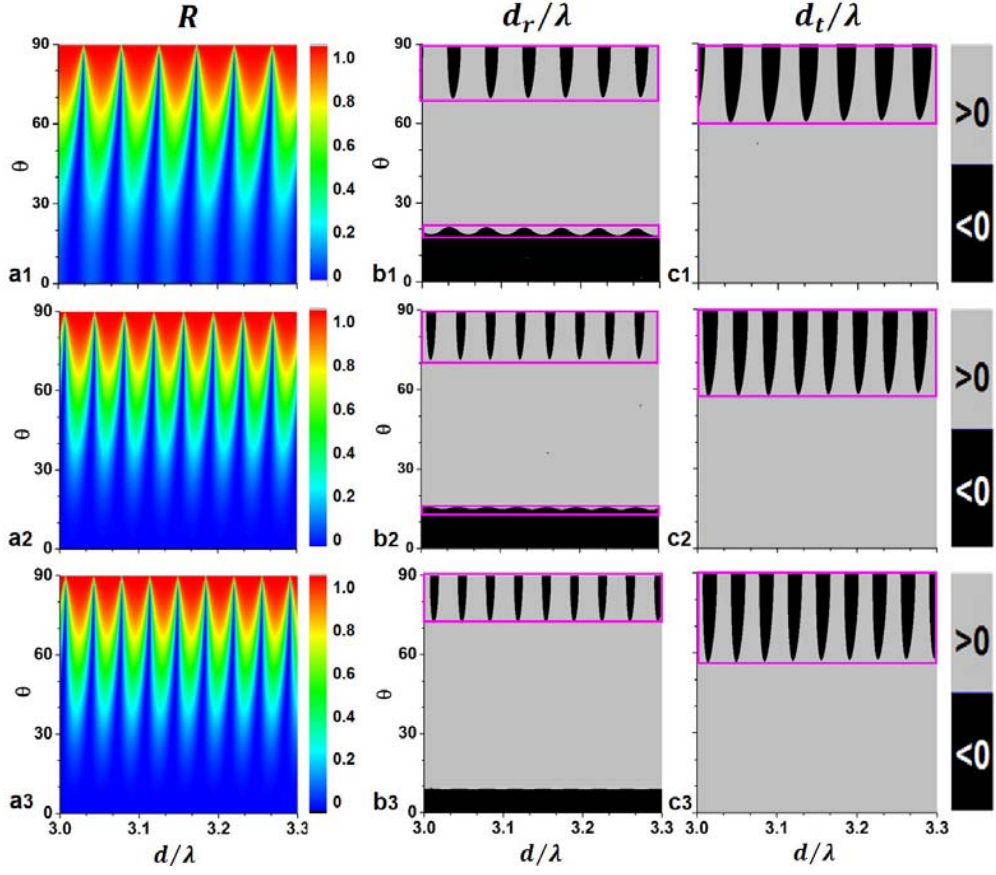

FIG. 3. (a1) Reflectivity, (b1) GH shift of reflected field $d_r/\lambda$ and (c1) GH shift of transmitted field $d_t/\lambda$ as functions of the slab thickness (expressed as $d/\lambda$) and the incident angle $\theta$ for $h_0 = 2780\text{Oe}$, $f = 9.5\text{GHz}$. (2), (3) are the same as (1) but for $f = 9.7\text{GHz}$ and $9.736\text{GHz}$, respectively.

The competition between $D(\theta)$ and $\Lambda(\theta,d)$ leads to the coexistence of two types of NRS at certain frequencies. Fig. 3 and Fig. 4 illustrate the variance of $|r|^2$, $d_r/\lambda$ and $d_t/\lambda$ with the incident angle $\theta$ and the slab thickness d. The magnetic field $h_0$ is set to be 2780Oe, at which the characteristic frequencies are given by $f_c = 9.749\text{GHz}$ and $f_r = 9.991\text{GHz}$. To one's interest, both reflectivity and the shifts show the periodicity with the change of slab-thickness (shown in Fig. 3). Two NRS regions are revealed in the sign-patterns of the lateral shifts for $f = 9.5\text{GHz}$, 9.7GHz and 9.736GHz in Fig. 3b and 3c, where region A extends from $\theta = 0$ to $\theta = \theta_A$ with only slight thickness dependence while region B for $\theta > \theta_B$ shows a periodic positive-to-negative transition of $d_r$ (and $d_t$ as well) with thickness varying.

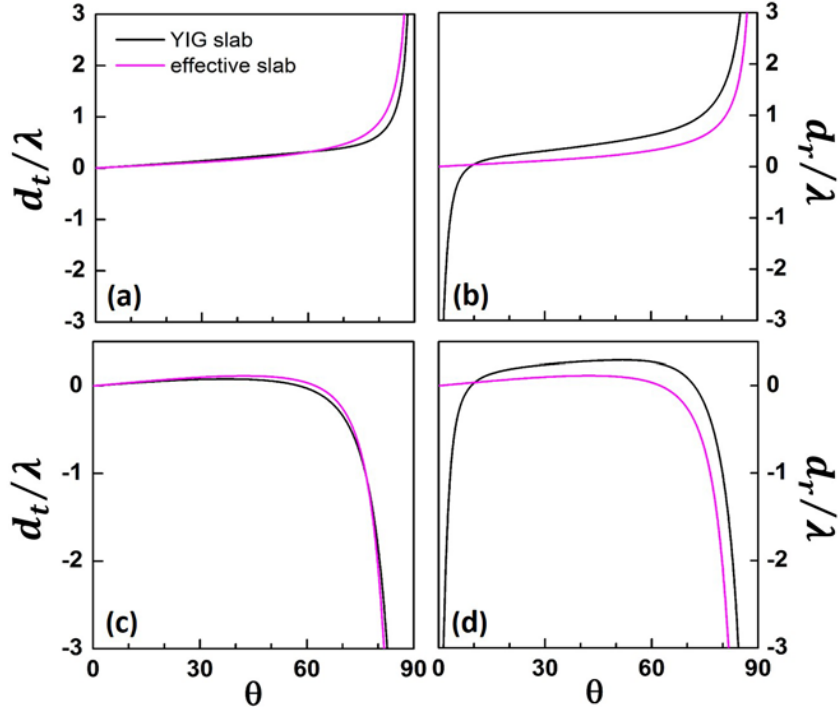

FIG. 4. (a), (c) GH shift of transmitted field $d_t/\lambda$ and (b), (d) GH shift of reflected field $d_r/\lambda$ vs the incident angle $\theta$ at two certain slab thicknesses: (a, b) $d = 3.03\lambda$, (c, d) $d = 3.047\lambda$ for both the YIG slab and the corresponding effective slab. The incidence frequency is $f = 9.736\text{GHz}$.

In Fig. 4(a) and (c), the curves of $d_t$ vs $\theta$ at a certain slab thickness for $f = 9.736\text{GHz}$ are presented for both the YIG slab and the corresponding effective slab. It is clearly seen that $\Lambda(\theta,d)$ can be well approximated by $\Lambda_{eff}(\theta,d)$, which accounts for the transition of $d_t$ with thickness at larger incident angles. For the reflected shift $d_r$ (Fig. 4(b) and (d)), $D(\theta)$ dominates the NRS region at smaller angles and makes a non-negligible correction to the NRS in region B, breaking the symmetry between $d_r$ and $d_t$ which is an important feature of GH shifts due to a non-MO slab[38].

**Two special cases.** Asymptotic behaviors of the GH shifts in two special cases of particular interest for applications can be obtained from the general formulas Eq. (24)-(30). The first case is at $\omega = \omega_r$, where total reflection occurs and the reflected shift is only determined by $D(\theta)$ even at oblique incidence. By expanding the function in terms of $\kappa \equiv k_y/k_0 = \sin\theta$ and keeping terms up to the second order, we have

$$D(\theta) = \frac{g}{k_0}\left[\left(\frac{1}{a_+}+\frac{1}{a_-}\right)+\left(\frac{2a_+b_+ - g^2}{2a_+^3}+\frac{2a_-b_- - g^2}{2a_-^3}\right)\kappa^2\right] \quad (31)$$

with $a_\pm = \mu_{eff} \pm n_{eff}$ and $b_\pm = \mu_{eff} \pm \frac{1}{n_{eff}}$.

Since $g/\mu_{eff} = \sqrt{H/(H+1)}$ at $\omega_r$, the asymptotic behavior of $d_r$ is given by

$$d_r\big|_{\omega=\omega_r} = d_r^{(n)}\left(1+\frac{H+2}{H+1}\kappa^2\right) \quad (32)$$

where $d_r^{(n)}$ is the reflected shift in Eq. (23) at normal incidence. The calculated results from Eq. (32) are illustrated in Fig. 5a in comparison with the numerical results for $h_0 = 1000\text{Oe}$, $2000\text{Oe}$ and $3000\text{Oe}$.

The second case is the transmitted shift accompanied by a 100% transmittivity when the slab thickness satisfies $k_{2x}d = m\pi (m \in \mathbb{Z})$. According to Eq. (24b), the transmitted shift can be written as

$$d_t = \frac{d\phi/dk_y}{1+\phi^2} \quad (33)$$

with

$$\phi = \frac{A_{0e}^2 + B_{0e}^2 + 2g^2 k_y^2}{A_{0e}^2 - B_{0e}^2}\tan(k_{2x}d) \quad (34)$$

when $k_{2x}d = m\pi$, we have $\phi = 0$ and

$$\frac{d\phi}{dk_y} = \left(-\frac{k_y}{k_{2x}}\right)\frac{A_{0e}^2 + B_{0e}^2 + 2g^2 k_y^2}{A_{0e}^2 - B_{0e}^2} \quad (35)$$

Also keeping the first two terms in the expression of $d_t$, we obtain the asymptotic behavior of $d_t$ in this case

$$d_t = D\kappa(1+\chi\kappa^2) \quad (36)$$

where the coefficients are given by

$$D = \frac{m\lambda_0}{2n_{eff}}\frac{\mu_{eff}+\varepsilon}{2\mu_{eff}\varepsilon} \quad (37)$$

and

$$\chi = \frac{n_{eff}^2 + 3}{2n_{eff}^2} - \frac{\mu_{eff}^2 + 1 - g^2}{\mu_{eff}^2 + n_{eff}^2} \quad (38)$$

The transmitted shift will vanish at $\omega = \omega_r$, because of the divergence of $\mu_{eff}$ at this frequency, and then rise with frequency decreasing. Fig. 5b illustrates the frequency-dependence of $d_t$ at a certain incident angle ($\theta = 30^o, 45^o$) for $h_0 = 3000 Oe$ when the slab thickness satisfies the total transmission condition. Good agreement is found between the approximated $d_t$ in Eq. (36) and the numerical results.

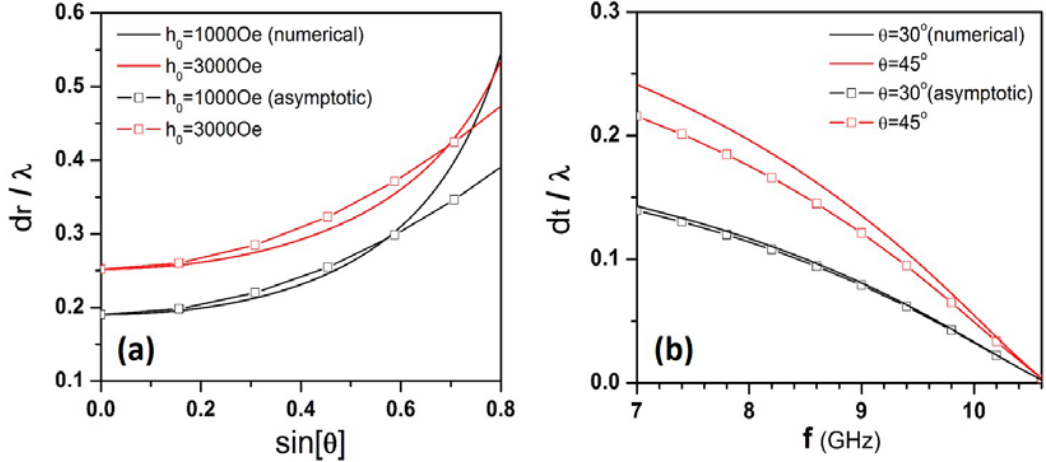

FIG. 5. (a) GH shift of reflected field $d_r / \lambda$ vs $\sin[\theta]$ for $h_0 = 1000, 3000 Oe$ at $f = f_r$. (b) GH shift of transmitted field $d_t / \lambda$ vs the frequency at two certain incident angles ($\theta = 30^o, 45^o$) for $h_0 = 3000 Oe$, $d = 10\pi / k_{2x}$. The solid lines indicate the numerical results and the square symbol lines correspond to the asymptotic behaviors calculated from Eq. (32) and (36).

**Numerical simulations**

To verify the above theoretical analysis, we performed a numerical simulation of a YIG slab illuminated by a Gaussian incident beam with the well-known finite-element analysis software COMSOL Multiphysics. The center of the incident beam arrived at the upper interface of the slab is located at the point $(0,0)$ and the half-width of the beam is $7.5\lambda$. The GH shifts can be directly obtained by comparing the field distributions of the incident beam and the reflected/transmitted beam at the relevant interfaces.

Note that the damping of YIG has been neglected in the analytic expressions. In our simulations, a more practical dispersion of YIG permeability will be adopted where the damping factor is set to be $\alpha = \gamma \times dH / 2\omega$, with $dH = 3 Oe$ [35]. The low damping ($\sim 10^{-4}$) implies that no significant absorption effects will occur except for frequencies near ferromagnetic resonance $f_0 = H f_m$. According to Eqs. (17), (18) and (20), the two characteristic frequencies for nonreciprocal GH shifts, $f_c$ and $f_r$, will not be close to $f_0$ unless the field $h_0$ is in the strong-field limit $h_0 \gg m_s$.

At normal incidence the analytical results predict that nonvanishing reflected shift occurs in both the transparent region ( $f < f_r$ ) and the opaque region ( $f > f_r$ ) as shown in Fig.2. The simulation results of the field distribution along the incident interface for both the incident beam and the reflected beam are given in Fig. 6. The parameters are chosen to be the same as those for the points A and B in Fig. 2b, namely $h_0$=2680Oe, $d$=0.3m, $f = 9.01$GHz (point A) or $f = 9.72$GHz (point B). The cases with (black solid lines) and without (blue solid lines) damping are both investigated. Table 1 gives the reflected shifts given by analytic expressions, simulations without damping and simulations with damping. It is shown that the damping has no significant effect on the shift, and the analytic predictions is in good agreement with the numerical results.

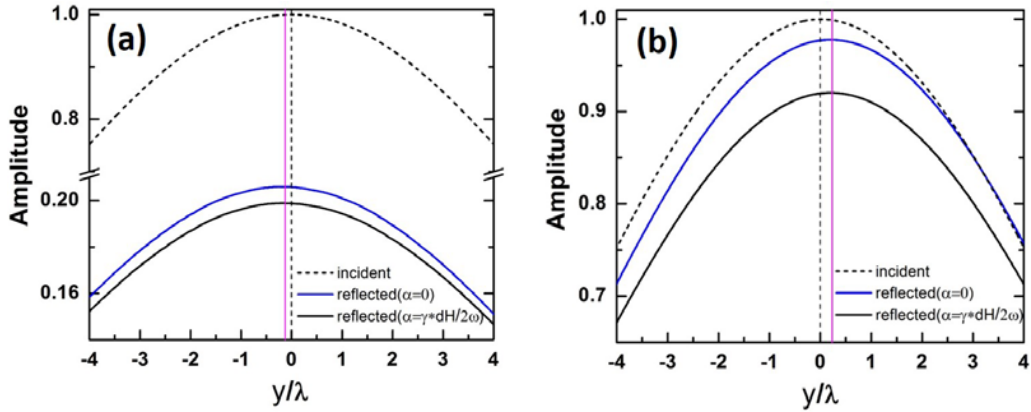

Fig. 6. The COMSOL simulation results for reflected shifts at normal incidence. (a) The distribution of electric field amplitude along the incident interface for $f = 9.01$GHz, $h_0 = 2680$Oe and $d = 0.3$m (corresponding to point A in Fig.2b); (b) The distribution of electric field amplitude along the incident interface for $f = 9.72$GHz $h_0 = 2680$Oe and $d = 0.3$m (point B in Fig.2b). The red lines indicate the analytical shift of each case.

Table 1: Comparisons between analytical and simulation results

|  |  | Analytical predictions | Simulations without damping | Simulations with damping |
|---|---|---|---|---|
| Normal incidence | $f$=9.01GHz | $d_r = -0.130\lambda$ | $d_r = -0.134\lambda$ | $d_r = -0.134\lambda$ |
|  | $f$=9.72GHz | $d_r = 0.229\lambda$ | $d_r = 0.221\lambda$ | $d_r = 0.221\lambda$ |
| Oblique incidence | Total reflection | $d_r = 0.443\lambda$ | $d_r = 0.454\lambda$ | $d_r = 0.454\lambda$ |
|  | Total transmission | $d_t = 0.241\lambda$ | $d_t = 0.245\lambda$ | $d_t = 0.245\lambda$ |

At oblique incidence both reflected and transmitted shifts may be observed at certain conditions. Fig.7 gives the simulated results when the incident angle is $45^o$ and the external magnetic field $h_0$ is $3000 Oe$. The frequency and the slab thickness are chosen to satisfy the conditions for total reflection ( $f = f_r$, Fig. 7a and 7c) and total transmission ( $k_{2x}d = m\pi$, Fig. 7b and 7d), respectively, since these cases are especially interesting for practical applications. Again both the cases with and without damping are investigated and compared with the analytical results as listed in Table 1. Trivial damping effects and good agreement between the analytical and simulation results are found, similar to those at normal incidence.

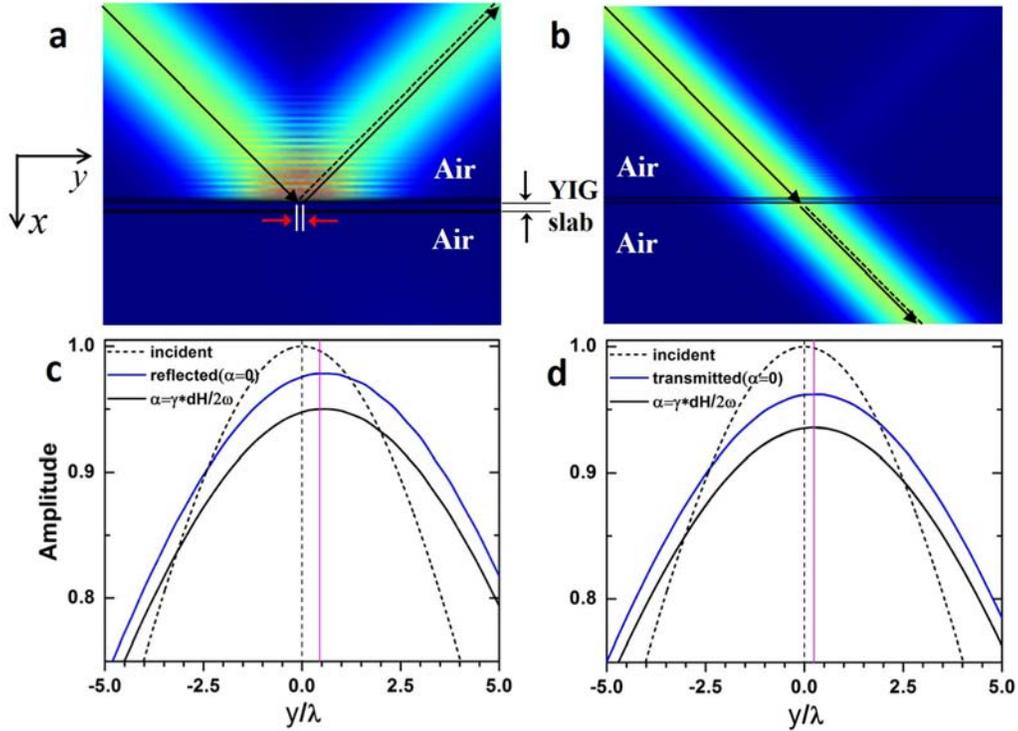

FIG. 7. Numerical simulations of GH shifts when the Gaussian beam is incident from air. (a) The field pattern for $h_0 = 3000 Oe$, $f = f_r$ with an incident angle of $45^o$. (b) The field pattern for $h_0 = 3000 Oe$, $f = 7 GHz$ and $d = 2\pi / k_{2x}$ with an incident angle of $45^o$. (c), (d) The distributions of field amplitudes along y direction near the interface between YIG and air, based on numerical results in (a) and (b). The red lines indicate the analytical shift of each case.

**Conclusions**

In this paper, we mainly investigate the lateral shifts of a TE wave both reflected and transmitted from a YIG slab theoretically. It is shown that the nonreciprocity effect caused by the MO material will result in a nonvanishing reflected shift at normal incidence. In the case of oblique incidence, this effect also leads to a slab-thickness-independent term of $d_r$ which breaks the symmetry between the reflected and transmitted shifts which is an important feature of GH shifts due to a non-MO slab. The asymptotic behaviors of the normal-incidence reflected shift are obtained in the vicinity of two characteristic frequencies ($\omega_r$ and $\omega_c$) corresponding to a minimum reflectivity and a total reflection, respectively. And the coexistence of two types of negative-reflected-shift (NRS) at oblique incidence is discussed. Numerical results show that the reversal of the sign of GH shifts can be realized by tuning the magnitude of external magnetic field $h_0$, adjusting the incident wave frequency $f$ or changing the thickness $d$ as well as the incident angle $\theta$. We also investigate two special cases for practical purposes: the reflected shift with a total reflection and the transmitted shift with a total transmission. Analytical expressions of the shifts in these two cases are obtained approximately, which are in good agreement with the results from numerical calculations.

Though nonreciprocal reflected shifts were also reported in antiferromagnetic $MnF_2$[16,17], our YIG-based study confirms the possibility of experimental demonstration of these effects in conventional ferrites at room temperatures. And the systematic analysis of both the reflected and the transmitted shifts due to a YIG slab offers a deeper insight into the role of magnetic field in tuning the shift sign, magnitude and types (reflected or transmitted).

**Methods**

**Theory and simulations.** The numerical simulation results shown in Fig. 6 and Fig. 7 were obtained using the finite element solver COMSOL Multiphysics. The scattering boundaries were set for four sides. Based on the numerical simulation, the curves of field amplitude in Fig. 6 were obtained by performing the line plot along y axis from $-4\lambda$ to $4\lambda$. Due to the interference effect, the field amplitudes are oscillating along x direction. The line plot is located at the first peak close to the interface between air and YIG. Meanwhile, we zoom in the line plot of $|E_z|$ enough to get the distance between its symmetric axis and y=0, which indicates the lateral shift $d_r$. The numerical results in Fig. 7 were obtained by the same technique.

**Acknowledgments**

This work was supported by the National Natural Science Foundation of China (Grant No. 11374223), the National Science of Jiangsu Province (Grant No. BK20161210), the Qing Lan project, "333" project (Grant No. BRA2015353), and PAPD of Jiangsu Higher Education Institutions.

**Author Contributions**

L.G. conceived the idea. W.Y. performed most theoretical and numerical calculations. W.Y. and H.S. analyzed the data. All authors joined discussion extensively and revised the manuscript before the submission.

**Additional Information**
**Competing financial interests:** The authors declare no competing financial interests.